\documentclass[a4paper,british,twocolumn]{svjour3}
\usepackage[T1]{fontenc}
\usepackage[latin9]{inputenc}
\usepackage{babel}
\usepackage{textcomp}
\usepackage{amsmath}
\usepackage{amssymb}
\usepackage{graphicx}
\usepackage[numbers]{natbib}
\usepackage[unicode=true]
 {hyperref}

\makeatletter

\pdfpageheight\paperheight
\pdfpagewidth\paperwidth

\makeatother

\begin{document}
\selectlanguage{british}
\newcommand{\mnras}{MNRAS}
\newcommand{\apjl}{ApJ}
\newcommand{\apj}{ApJL}
\newcommand{\aj}{AJ}
\newcommand{\aap}{A\&A}
\newcommand{\procspie}{SPIE}
\newcommand{\mdash}{-}
\journalname{Publications of the Astronomical Society of the Pacific}
\institute{
Millenium Institute of Astrophysics, Vicu\~{n}a. MacKenna 4860, 7820436 Macul, Santiago, Chile
\\
Pontificia Universidad Católica de Chile, Instituto de Astrofísica, Casilla 306, Santiago 22, Chile
\\
Excellence Cluster Universe, Boltzmannstr. 2, D-85748, Garching, Germany
\\
\email{johannes.buchner.acad@gmx.com}
}
%@arxiver{outimg_noise0p2.pdf,plotscaling.pdf,plotcontour_5.pdf}

\title{Collaborative Nested Sampling: Big Data vs. complex physical models}

\author{Johannes Buchner}

\date{~}
\maketitle
\begin{abstract}
The data torrent unleashed by current and upcoming astronomical surveys
demands scalable analysis methods. Many machine learning approaches
scale well, but separating the instrument measurement from the physical
effects of interest, dealing with variable errors, and deriving parameter
uncertainties is often an after-thought. Classic forward-folding analyses
with Markov Chain Monte Carlo or Nested Sampling enable parameter
estimation and model comparison, even for complex and slow-to-evaluate
physical models. However, these approaches require independent runs
for each data set, implying an unfeasible number of model evaluations
in the Big Data regime. Here I present a new algorithm, collaborative
nested sampling, for deriving parameter probability distributions
for each observation. Importantly, the number of physical model evaluations
scales sub-linearly with the number of data sets, and no assumptions
about homogeneous errors, Gaussianity, the form of the model or heterogeneity/completeness
of the observations need to be made. Collaborative nested sampling
has immediate application in speeding up analyses of large surveys,
integral-field-unit observations, and Monte Carlo simulations.
\end{abstract}

\keywords{Nested sampling, Big Data, Bayesian inference}

\section{Introduction}

%@arxiver{outimg_noise0p2.pdf,plotscaling.pdf,plotcontour_5.pdf}

Big Data has arrived in astronomy \citep{Feigelson2012,Zhang2015a,Mickaelian2016,Kremer2017}.
In the previous century it was common to analyse a few dozen objects
in detail. For instance, one would use Markov Chain Monte Carlo to
forward fold a physical model and constrain its parameters. This would
be repeated for each member of the sample. However, current and upcoming
instruments provide a wealth of data ($\sim$ millions of independent
sources) where it becomes computationally difficult to follow the
same approach, even though it is embarrassingly parallel. Currently,
much effort is put into studying and applying machine learning algorithms
such as (deep learning) neural networks or random forests for the
analysis of massive datasets. This can work well if the measurement
errors are homogeneous, but typically these methods make it difficult
to insert existing physical knowledge into the analysis, to deal with
variable errors and missing data points, and generally to separate
the instrument measurement process from the physical effects of interest.
Furthermore, we would like to derive probability density distributions
of physical parameters for each object, and do model comparison between
physical effects/sources classes.

In this work I show how nested sampling can be used to analyse $N$
data sets simultaneously. The key insight is that nested sampling
allows effective sharing of evaluation points across data sets, requiring
much fewer model evaluations than if the $N$ data sets were analysed
individually. I only assume that the model can be split into two components:
a slow-to-evaluate physical model which performs a prediction into
observable space, and a fast-to-compute comparison to the individual
data sets (e.g. the likelihood of a probability distribution). Otherwise,
the user is free to chose arbitrary physical models and likelihoods.
§\ref{sec:Results} presents a line fitting of a hypothetical many-object
spectroscopic survey as a toy example; §\ref{sec:ifuresults} constrains
the properties of stellar populations in a real imaging-spectroscopy
observation.

\section{Methodology}

\subsection{Introduction to Classic Nested Sampling}

\begin{figure}
\includegraphics[viewport=0bp 10bp 369bp 160bp,clip,width=1\columnwidth]{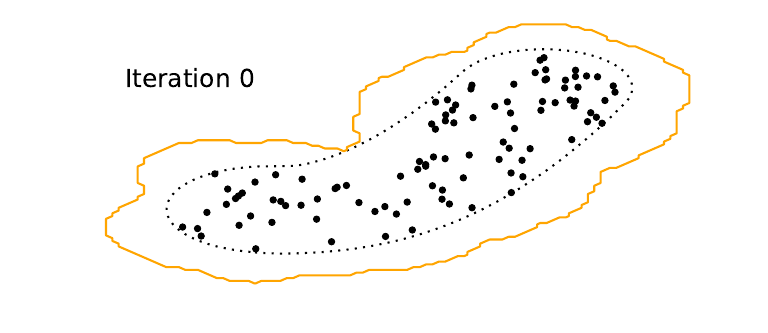}

\includegraphics[viewport=0bp 30bp 369bp 140bp,clip,width=1\columnwidth]{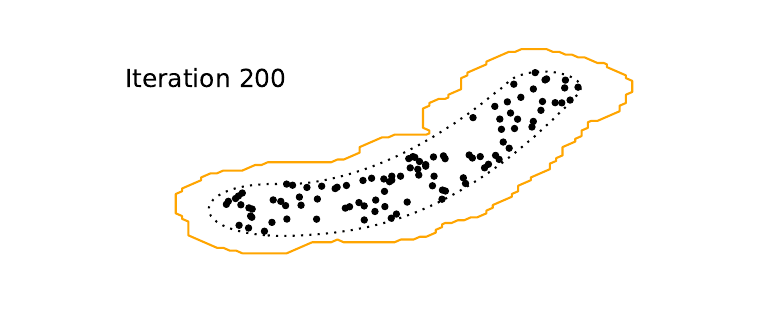}

\includegraphics[viewport=0bp 40bp 369bp 130bp,clip,width=1\columnwidth]{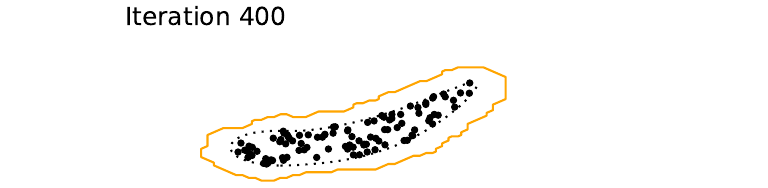}

\includegraphics[viewport=0bp 50bp 369bp 130bp,clip,width=1\columnwidth]{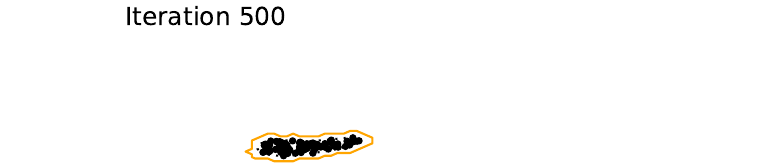}

\caption{\label{fig:IllustrationNS}Illustration of nested sampling. At a given
iteration of the nested sampling algorithm, the live points (black)
trace out the current likelihood constraint, a region (dashed) which
is unknown. The \textsc{RadFriends} algorithm conservatively reconstructs
the region (orange) by including everything within a certain, adaptively
chosen radius of the current live points. Between iterations, the
likelihood contour is elevated, making the sampled volume smaller
and smaller. \textsc{MultiNest} works similarly, but clusters point
into ellipsoids.}
\end{figure}
Nested sampling \citep{Skilling2004} is a global parameter space
exploration algorithm, which zooms in from the entire volume towards
the best-fit models by steadily increasing the likelihood threshold.
In the process it produces parameter posterior probability distributions
and computes the integral over the parameter space. Assume that the
parameter space is a $k$-dimensional cube. A number of live points
$N_{{\rm live}}$ are randomly\footnote{In general, following the prior. For most problems one can uniformly
sample with appropriate stretching of the parameter space under the
inverse cumulative of the prior distributions \citep[see §5.1 in][]{Feroz2009}.} placed in the parameter space. Their likelihood is evaluated. Each
point represents $1/N_{{\rm live}}$ of the entire volume. The live
point with the lowest likelihood $L_{{\rm min}}$ is then removed,
implying the removal of space with likelihood below $L_{{\rm min}}$
and shrinkage of the volume to $1-\exp\left(-1/N_{{\rm live}}\right)$,
on average. A new random live point is drawn, with the requirement
that its likelihood must be above $L_{{\rm min}}$. This replacement
procedure is iterated, shrinking the volume exponentially. Each removed
(``dead'') point and its likelihood $L_{i}$ is stored. The integral
over the parameter space can then be approximated by $Z=\sum_{i}L_{i}\times w_{i}$,
where $w_{i}$ is the removed volume at the iteration. At a late stage
in the algorithm the volume probed is tiny and the likelihood $L_{i}$
increase is negligible, so that the weights $L_{i}\times w_{i}$ of
the remaining live points becomes small. Then the iterative procedure
can be stopped (the algorithm converged). The posterior probability
distribution of the parameters is approximated as importance samples
of weight $L_{i}\times w_{i}$ at the dead point locations, and can
be resampled into a set of points with equal weights, for posterior
analyses similar to those with Markov Chains. More details on the
convergence and error estimates can be found in \citep{skilling2009nested}.

Efficient general solutions exist for drawing a new point above a
likelihood threshold in low dimensions ($n_{{\rm dim}}<20$). The
idea is to draw only in the neighbourhood of the current live points,
which already fulfill the likelihood threshold. The best-known algorithm
in astrophysics and cosmology is \textsc{multinest} \citep{Shaw2007,Feroz2009}.
There, the contours traced out by the points are clustered into ellipses,
and new points drawn from the ellipses. To avoid accidentally cutting
away too much of the parameter space, the tightest-fitting ellipses
are enlarged by an empirical (problem-specific) factor. Another algorithm
is \textsc{RadFriends} \citep{buchner2014statistical}, which defines
the neighbourhood as all points within a radius $r$ of an existing
live point. By leaving out randomly a portion of the live points,
and determining their distance to the remaining live points, the largest
nearest-neighbour radius $r$ is determined. The worst-case analysis
through bootstrapping cross-validation over multiple rounds makes
\textsc{RadFriends} robust, independent of contour shapes and free
of tuning parameters. Figure~\ref{fig:IllustrationNS} illustrates
the generated regions. \textsc{RadFriends} is efficient if one chooses
a standardised euclidean metric (i.e. normalise by the standard deviation
of the live points along each axis). The extension to nested sampling
proposed in this paper works with any constrained drawing method.

\subsection{Simplified description of the idea}

\begin{figure}
\includegraphics[viewport=45bp 0bp 369bp 160bp,clip,width=1\columnwidth]{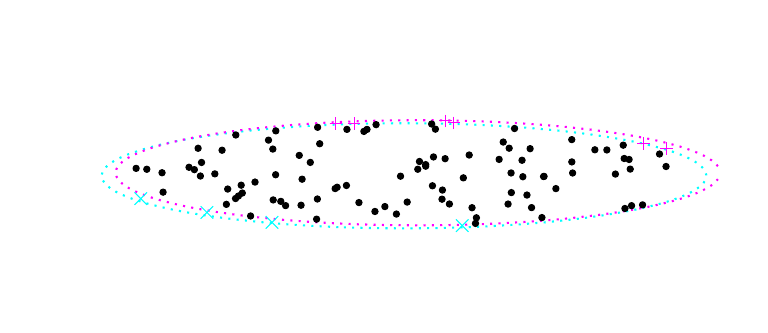}

\caption{\label{fig:illustration-pointsharing}The analysis of two similar
data sets yields at the same iteration similar likelihood contours
(the two dotted ellipses). In the presented algorithm a large fraction
of live points are shared across data sets (black points), which reduces
the number of model evaluations. The differences (cyan crosses and
magenta pluses) requiring additional draws.}
\end{figure}

Consider two independent nested sampling runs on different data sets,
but initialised to the same random number generator state. Initially
points are generated from across the entire parameter space, typically
giving bad fits. If the data sets are somewhat similar, the phase
of zooming to the relevant parameter space will be the same for the
two runs. Importantly, while the exact likelihood value will be different
for the same point, the ordering of the points will be similar. In
other words, for both, the worst-fitting point to be removed is likely
the same. The next key insight is that new points can be drawn efficiently
from a contour which is the union of the likelihood contours from
both runs. Ideally, the point can be accepted by both runs, keeping
the runs similar (black points in Figure~\ref{fig:illustration-pointsharing}).
When a point is shared, the (slow) predicting model has to be only
evaluated once, speeding up the run. The model prediction is then
compared against the data to produce a likelihood for each data set,
an operation which I presume to be fast, e.g., when computing 
\begin{equation}
{\cal L}_{j}=-\sum_{i}(x_{ij}-m_{i})^{2}/(2\sigma_{ij}^{2})\label{eq:chisquare}
\end{equation}
 where $m_{i}$, $x_{ij}$ and $\sigma_{ij}$ are the predictions,
measurements and errors in data space respectively for data set $j$.

What if the point can be accepted by only one run? It cannot simply
be rejected or accepted in both, otherwise the uniform sampling property
of nested sampling is broken. Instead, accepted points are stored
in queues, one for each run/data set. Once both runs have a non-empty
queue, the first accepted point is removed from each queue and replaces
the dead point of each data set. Joint sampling also helps even if
a point is not useful right away. If a point was only accepted by
run A, but the following point is accepted by both runs, the second
point becomes a live point immediately for run B, but can later also
become a live point for run A (if it suffices the likelihood threshold
at that later iteration). This technique allows sustained sharing
of points, decreasing the number of unique live points and increasing
the speed-up.

At a later point in the algorithm, the contours may significantly
diverge and not share any live points. This is because the best-fit
parameters of data sets will differ. Then, nested sampling runs can
continue as in the classic case, without speed-up, falling back to
a linear scaling. This happens earlier, the more different the data
sets are. The run is longer for data sets with high signal-to-noise,
making the algorithm most efficient when most observations are near
the detection limit. This is typically the case in surveys as a consequence
of powerlaw distributions.

\subsection{Collaborative nested sampling}

I now describe the collaborative nested sampling algorithm. A proof-of-concept
reference implementation is available at \href{https://github.com/JohannesBuchner/massivedatans/}{https://github.com/JohannesBuchner/massivedatans/}.
The algorithm components are the nested sampling integrator, the constrained
sampler and the likelihood function, as in classic nested sampling,
except that works on $N$ data sets simultaneously, with $N$ a large
number. The constrained sampler behaves substantially different in
this algorithm.

\subsubsection{Likelihood Function}

The likelihood function receives a single parameter vector, and information
which data sets to consider. It calls the physical model with the
parameter vector to compute into a prediction into data space. The
physical model may perform complex and slow numerical computations/simulations
at this point. Finally the prediction is compared with the individual
data sets to produce a likelihood for each considered data set. The
likelihood at this point can be Gaussian (eq.~\ref{eq:chisquare}),
Poisson, a red noise process, or any other probability distribution
appropriate for the instrument. In any case, this computation must
be fast compared with producing the model predictions to receive any
performance gains.

\subsubsection{Nested Sampling Integrator}

The integrator deals with each run individually just as in standard
nested sampling. It keeps track of the remaining volume at the current
iteration, and storing the live points and their weights for each
data set individually. It calls the constrained sampler (see below),
which holds the live points, to receive the next dead point (for all
data sets simultaneously). The integrator must also test for convergence,
and advance further only those runs which have not yet converged.
Here I use the standard criterion that the nested sampling error is
$\delta Z<0.5$ (from last equation in \citealp{skilling2009nested}).
Once all runs have terminated, corresponding to each data set the
integral estimates $Z$ and posterior samples are returned, giving
the user the same output as e.g., a \textsc{multinest} analysis.

\subsubsection{Constrained Sampler}

\begin{figure}
\includegraphics[width=1\columnwidth]{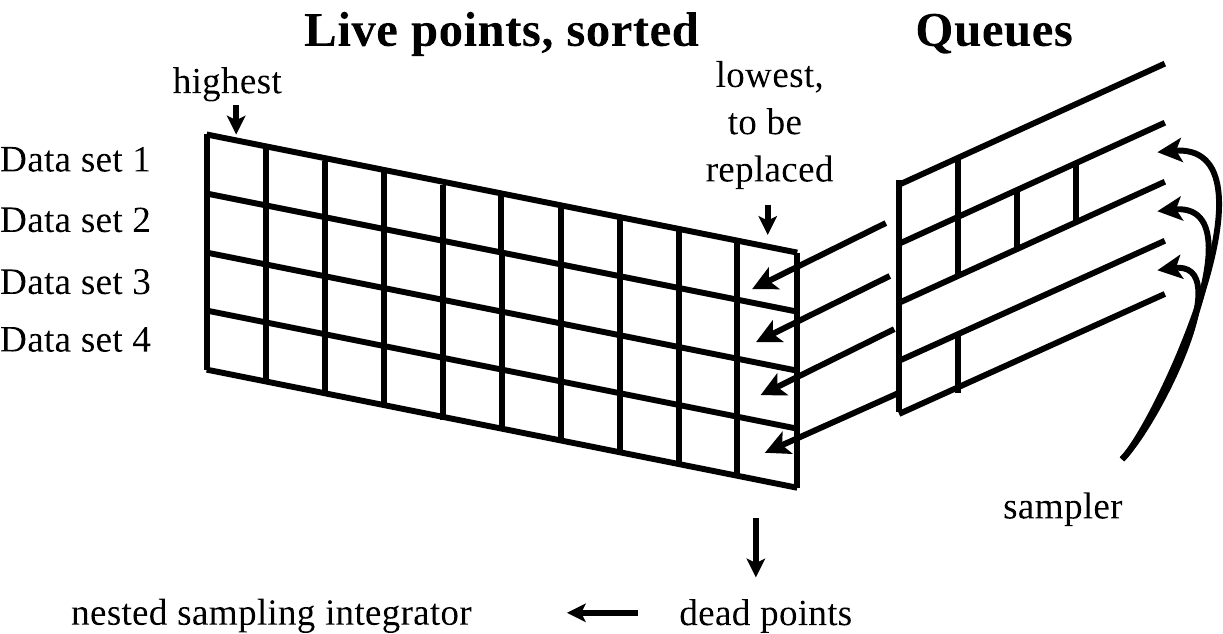}

\caption{\label{fig:queues}To replace the least likely live point, new points
are sampled and placed in queues if they have a high enough likelihood.
Once every data set has a non-empty queue, the lowest points are pushed
out and stored as dead points by the integrator. In this illustration,
$N=4$ data sets are sampled with $N_{{\rm live}}=10$ live points.}

\end{figure}

The sampler initially draws $N_{{\rm live}}$ live points and stores
their likelihoods in an array of size $N\times N_{{\rm live}}$. Sequential
IDs are assigned to live points and the mapping between live point
IDs and data sets ($N\times N_{{\rm live}}$ indices) is stored. The
integrator informs the sampler when it should remove the lowest likelihood
point and replace it. The integrator also informs the sampler when
some data sets have finished and can be discarded from further consideration,
in which case the sampler works as if they had never participated.

The main task of the constrained sampler is to do joint draws under
likelihood constraint $L>L_{{\rm min}}$ to replace the lowest likelihood
point in each of the $d$ data sets. For this, $d$ initially empty
queues are introduced (see Figure~\ref{fig:queues}). First, it is
attempted to draw from the joint contour over all data sets (\emph{superset
draw}), i.e. letting \textsc{RadFriends} define a region based on
the all unique live points. From this region a point is drawn which
has $L>L_{{\rm min}}$ for at least one data set. Some will accept
and the corresponding queues are filled. If this fails to fill all
queues after several (e.g. 10) attempts, a \emph{focussed draw} is
done. In that case, only the data sets with empty queues are considered,
the region is constructed from their live points, and the likelihood
only evaluated for these data sets. For example, in the illustration
of Figure~\ref{fig:queues}, only Data Set 3 would be considered.
Once all queues have at least one entry, nested sampling can advance:
For each data set, the first queue entry is removed and replaces the
dead live point. In Figure~\ref{fig:queues} this is illustrated
by the queues pushing out the lowest live points. These dead points
are returned to the integrator.

Storing queue entries is only useful if they can replace live points
in future nested sampling iterations. To be accepted into the queue
at position $j$, it must have a likelihood higher than $j$ points
from the runs live points and existing entries of the queue. In other
words, the first entry must merely beat a single existing live point,
the second entry must beat both a live point and either another live
point or the first queue entry (which will become a live point in
the next iteration).

\subsubsection{Data Set Clustering}

\begin{figure}
\includegraphics[width=1\columnwidth]{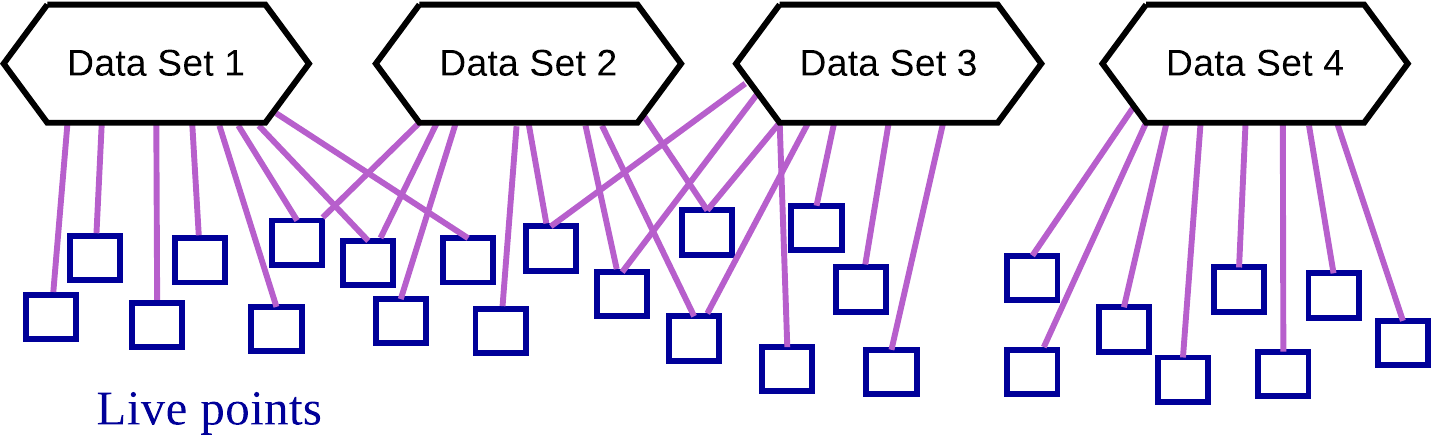}

\caption{\label{fig:graph}Association of live point objects with data sets.
In this illustration, some live points are shared between the group
of Data Sets 1, 2 and 3; these form a connected subgraph. Data Set
4 has separate live points and can be treated independently. In this
illustration, $N_{\mathrm{live}}=8$ and $N=4$, but there are only
$26$ unique live points.}
\end{figure}

It can occur that between two groups of data sets the live points
are not shared any more, i.e. the live point sets are disjoint (see
Figure~\ref{fig:graph}). For example, one may have a dichotomy between
broad and narrow line objects, and the contours identify some of the
data sets in the former, some in the latter class. Distinct groups
caused by diverging likelihood contours are interesting aspect of
the exploration: It defines a data set similarity through the constraints
in parameter space, based on the likelihood ordering unique to nested
sampling. This is different to clustering data sets in data space,
which can be non-trivial for varying errors and completeness, and
clustering in parameter space could scale poorly with model dimensionality.
In practice, diverging live points groups can be identified by finding
connected subsets in a graph. As illustrated in Figure~\ref{fig:graph},
the necessary graph can be constructed with nodes corresponding to
the data sets, nodes corresponding to the live points, and connecting
the graph according to the current live point statuses. Algorithms
for identifying connected subsets of graphs are well-known. These
data-set groups can be processed independently, avoiding multi-modal
contours. In the numerical examples shown in this work, this however
does not yield substantial speed-ups.

\section{Toy Application: Single-line fitting}

\label{sec:Results}

\begin{figure}
\includegraphics[width=1\columnwidth]{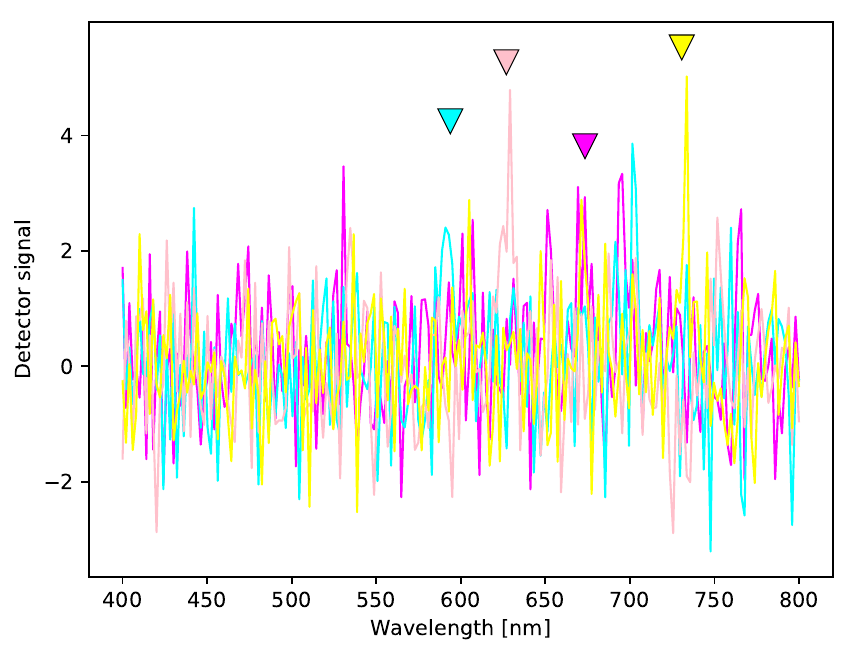}

\caption{\label{fig:spectra}Simulated noisy data. The location, width and
amplitude of a single line is sought in Gaussian noise for the illustrative
problem. The true line locations of the four spectra are indicated
by triangles. The cyan data set shows a random fluctuation at $700{\rm nm}$.}
\end{figure}
A simple toy example problem illustrates the use and scaling of the
algorithm. Lets consider a spectroscopic survey which collected $N$
spectra in the $400-800{\rm nm}$ wavelength range. We look for a
Gaussian line at $654{\rm nm}$ rest frame (but randomly shifted)
with standard deviation of $0.5{\rm nm}$. The amplitudes vary with
a powerlaw distribution with index $3$, with a signal-to-noise ratio
of at least two. I generate a large random data set and analyse the
first $N$ data sets simultaneously to understand the scaling of the
algorithm, with $N=1$ to $N=10^{4}$. Figure~\ref{fig:spectra}
presents some high and low signal-to-noise examples of the simulated
data set.

The parameter space of the analysis has three dimensions: The amplitude,
width and location of a single Gaussian line, with log-uniform/log-uniform/uniform
priors from $10^{0-2}$, $0.15-15{\rm nm}$ and $600-1000{\rm nm}$
respectively. The Gaussian line is our ``slow-to-compute'' physical
model. The likelihood function is as in equation~\ref{eq:chisquare}.
A more elaborate example would include physical modelling of an ionised
outflow emitting multiple lines with Doppler broadening and red detector
noise, without necessitating any modification of the presented algorithm.

\begin{figure}
\includegraphics[width=1\columnwidth]{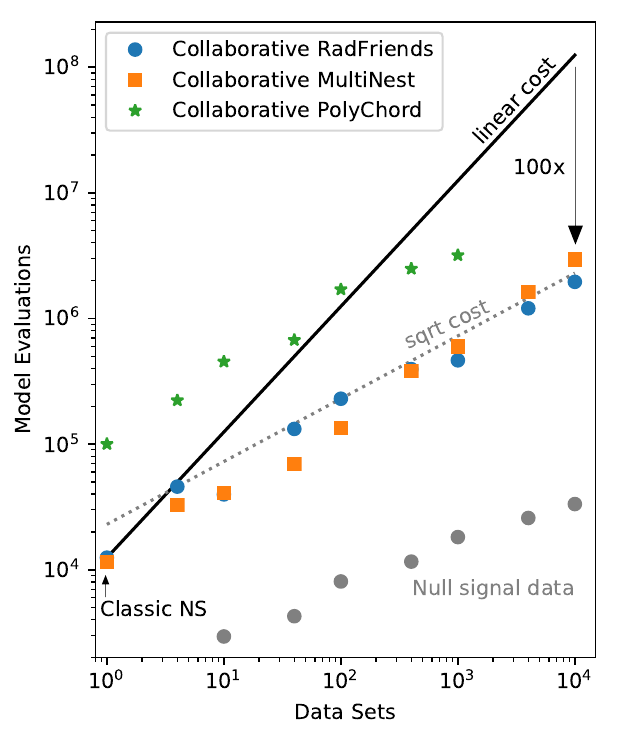}

\caption{\label{fig:cost}Number of model evaluations of collaborative nested
sampling applied to \textsc{RadFriends}, multi-ellipsoidal sampling
(\textsc{MultiNest}) and whitened slice sampling (\textsc{PolyChord}).
A naive approach of independent nested sampling analyses would have
a linear scaling (black line). The algorithm scales substantially
better, similar to $O(\sqrt{N})$ in the considered problem, giving
a 100x speed-up when analysing 10,000 data sets. Analysing Monte Carlo
simulated data without signal (gray points) is also faster.}
\end{figure}

Figure~\ref{fig:cost} shows the number of model evaluations necessary
for analysing $N$ data sets. We implemented our nested sampling variant
on top of three constrained drawing methods, \textsc{RadFriends} \citep{buchner2014statistical},
multi-ellipsoidal sampling (\textsc{MultiNest}, \citep{Shaw2007,Feroz2009})
and eigenvector slice sampling (\textsc{PolyChord}, \citealp{Handley2015a},
here for simplicity implemented without clustering). The black line
shows the baseline linear scaling $O(N)$, i.e. analysing the data
sets individually one-by-one. The algorithm scales much better, close
to $O(\sqrt{N})$. For instance, it takes only $100$ times more model
evaluations to analyse $10,000$ observations than a single observations,
a $100$-fold speedup.

\begin{figure}
\includegraphics[width=1\columnwidth]{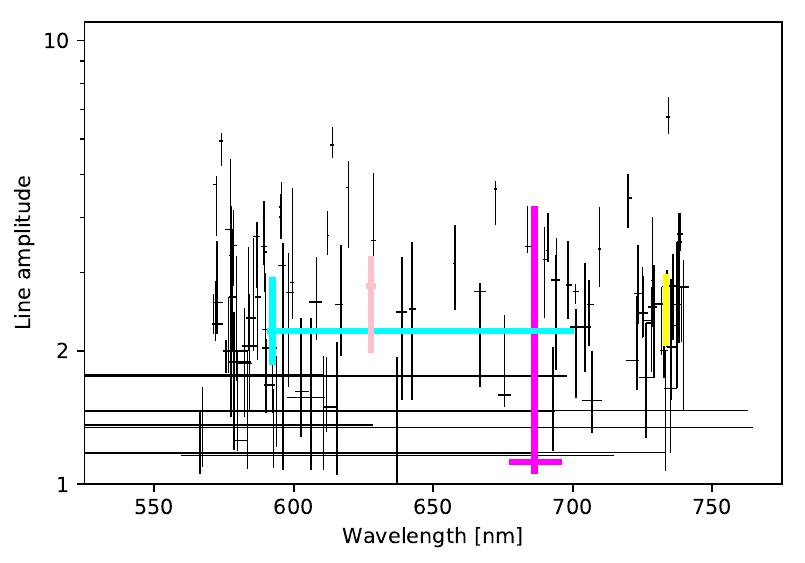}

\caption{\label{fig:post}Parameter posterior constraints. Each error bar shows
a simulated data set; the four examples from Figure~\ref{fig:spectra}
are shown in the same colours. The pink and yellow data sets have
been well-detected and characterized, while the magenta line has larger
uncertainties. The cyan constraints cover two solutions (see Figure~\ref{fig:spectra}).
Error bars are centred at the median of the posteriors with the line
lengths reflecting the 1-sigma equivalent quantiles. }
\end{figure}

\begin{figure}
\includegraphics[width=1\columnwidth]{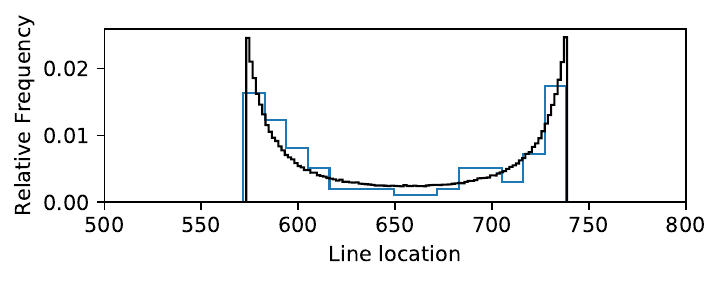}

\caption{\label{fig:postz}Line location distribution for objects where the
line was well-constrained (blue) compared to the input distribution
(black).}
\end{figure}

We can now plot the posterior distributions of the found line locations.
Figure~\ref{fig:post} demonstrates the wide variety of uncertainties.
The spectra of Figure~\ref{fig:spectra} are shown in the same colours.
For many, the line could be identified and characterised with small
uncertainties (yellow, pink, black), for others, the method remains
unsure (cyan, magenta). Figure~\ref{fig:postz} shows that the input
redshift distribution is correctly recovered.

\begin{figure}
\includegraphics[width=1\columnwidth]{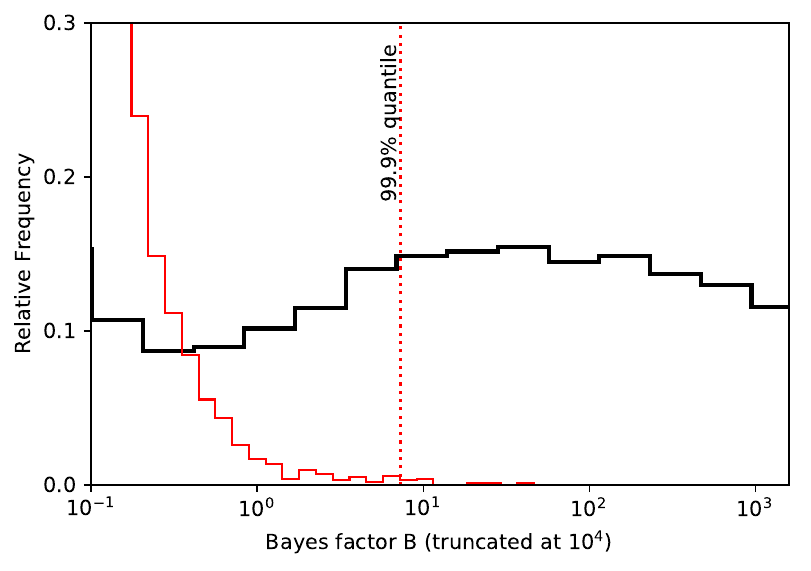}

\caption{\label{fig:postZ}Bayes factors between the single-line model and
a no-line model. The black histogram shows Bayes factors from analysing
the test data set. The red histogram shows Bayes factors from noise-only
data. Because the latter has very few values above $B\gtrapprox10$,
a line can be claimed detected beyond that threshold with a low false
positive fraction.}
\end{figure}

After parameter estimation we can consider model comparison: is the
line significantly detected? For this, lets consider the Bayes factor,
$B=Z_{1}/Z_{0}$, where $Z_{1}$ is the integral computed by nested
sampling under the single-line model, and $Z_{0}$ is the same for
the null hypothesis (no line). The latter can be analytically computed
as $\ln Z_{0}=-\frac{1}{2}\left[\sum(x_{i}/\sigma_{i})^{2}+\ln2\pi\sigma_{i}^{2}\right]$.
Figure~\ref{fig:postZ} shows in black the derived Bayes factors.
To define a lower threshold for significant detections, I Monte Carlo
simulate a dataset with $N=10^{4}$ spectra without signal, and derive
$Z_{1}$ values. This can be done rapidly with the presented algorithm.
The red histogram in Figure~\ref{fig:postZ} shows the resulting
Bayes factors. The $99.9\%$ quantile of $B$-values in this signal-free
data set is $B\approx10$. Therefore, in the ``real'' data, those
with a Bayes factor $B>10$ can be securely claimed to have a line,
with a small fraction of false detection ($p<0.001$).

\section{Application to Imaging Spectroscopy\label{sec:ifuresults}}

\begin{figure*}
\includegraphics[width=1\textwidth]{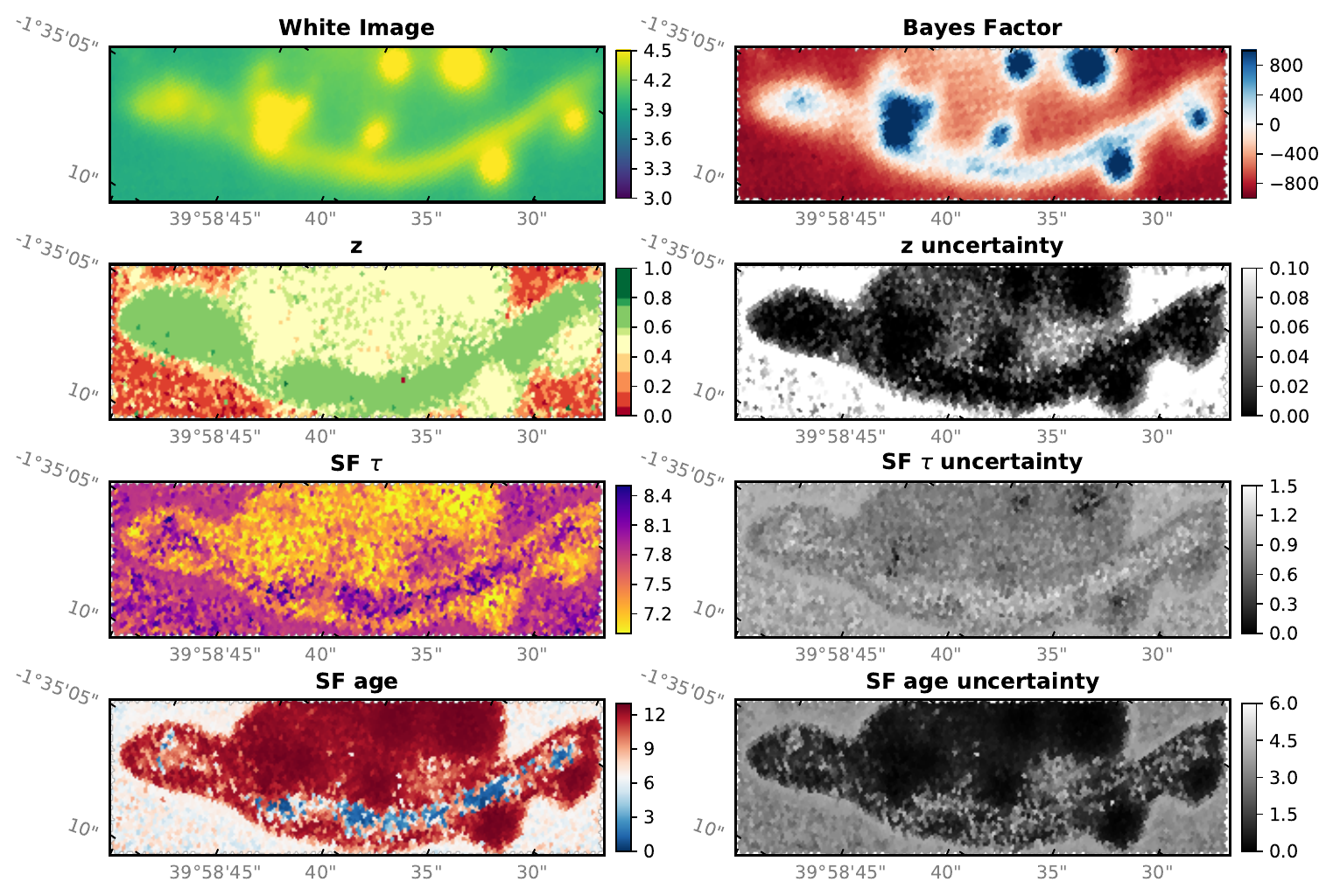}

\caption{\label{fig:ifuresults}MUSE IFU data analysed with Collaborative Nested
Sampling. \emph{Top left panel}: White image of the input data. Several
blobs and a extended arc is visible. \emph{Top right panel}: Bayes
factor $B$ comparing the single-stellar population model to a no-signal
model. In the arc and blobs, $\log B>0$. The remaining panels present
posterior parameters (\emph{left}) and uncertainties (\emph{right}).
Redshift $z$ is well-constrained across the arc ($z\approx0.6$)
and the closer blobs ($z\approx0.4$). While the blobs had a brief
star formation episode ($\log\tau/\mathrm{yr}\approx7$) long ago
(age > $10^{10}$ years), the arc shows evidence for recent star formation
(younger age, slower decay $\tau$).}

\end{figure*}

Finally, I apply collaborative nested sampling to a real-world data
set. Integral-field unit (IFU) observations, where many spectra are
taken in proximity on the sky are ideal for applying the algorithm.
The MUSE spectrograph (1arcmin\texttwosuperior{} field of view, wavelength
range 480-930nm; \citep{Bacon2010}) observed the Abell~370 galaxy
cluster in November 2014 (PI: Richard) for one hour\footnote{Additional observations have been made since then, however here the
demonstration is intended to show information extraction in the low-signal
regime.}. Following e.g., \citep{Lagattuta2017} and \citep{Patricio2018},
standard data reduction procedures and sky line subtraction (\citep{Soto2016ZAP})
were used, and the errors increased by 20\% of the data value to account
for model inaccuracies. The chosen region in the sky (A370-sys1 in
\citep{Patricio2018}) includes several galaxies, some of which heavily
distorted by strong lensing. Its white image (sum across the spectrum)
is shown in the top-left panel of Figure~\ref{fig:ifuresults}. The
169 arcsec\texttwosuperior{} sized region is covered by 4223 fibers,
each of which providing a spectrum with measured intensity and error.

A simple stellar population is used to model the spectra. The classic
Bruzual \& Charlot model stellar spectra \citep{Bruzual2003} are
weighed by an exponential star formation (decay timescale $\tau$)
at time $t$ Gyrs in the past. Additionally, dust extinction is allowed
through a Calzetti law\citep{Calzetti2000} and the model is redshifted.
To avoid a degeneracy with star formation age, I assume solar metallicity.
This model thus has four parameters: redshift $z$ (0-1), star formation
age $t$ (0-13~Gyr), star formation decay time $\tau$ ($10^{6-9}$
Gyr) and extinction $E(B-V)$ (0-1). Uniform priors are applied on
$z$, $t$, $\log\tau$ and $E(B-V)$.

A Gaussian likelihood compares the model spectrum $M_{i}$ against
the measurements $\mu_{i}$ and errors $\sigma_{i}$. To avoid having
the model normalisation $s$ as a fitting parameter, it is marginalised
over, by setting 
\begin{equation}
s=\sum_{i}\left(\mu_{i}M_{i}\sigma_{i}^{-2}\right)/\sum_{i}\left(M_{i}^{2}\sigma_{i}^{-2}\right)
\end{equation}
 and neglecting constants in the likelihood (see \citep{Arnouts1999}
for more details):

\begin{equation}
\log L=-\frac{1}{2}\sum\left[(\mu_{i}-s\cdot M_{i})/\sigma_{i}\right]^{2}
\end{equation}

With the model, likelihood and data defined, I apply collaborative
nested sampling with multi-ellipsoidal sampling and derive posterior
parameter distribution at each spaxel. Evidence values are also obtained.
Similar to the previous section, Bayes factors are computed and shown
in the top right panel in Figure~\ref{fig:ifuresults} at each spaxel.
The red areas indicate where the data prefer no input signal over
the stellar model. 

The second row of Figure~\ref{fig:ifuresults} shows the derived
redshift. Uncertainties (right panel) are extremely small (typically
0.001) over most of the image. Two solutions are visible in the left
panel: The arc (green) is at distinctly higher redshifts than the
blobs (yellow). Extended emission at the same redshift as the blobs
is detected.

The third and forth row present the star formation properties in each
pixel. The arc shows evidence for recent star formation. While uncertainties
on the decay parameter are not small, generally the arc has a longer
($10^{8}\,\mathrm{Gyr}$) star formation episode than the blobs ($10^{7}\,\mathrm{Gyr}$).
The extinction is constant over image and shows small values ($<0.1$;
not shown). The model used here is overly simple and I do not interpret
the physical meaning in great detail (see \citep{Patricio2018} instead).
In particular, the relation between metallicity and age should be
explored further. \citep{Patricio2018} however derived similar values
e.g., for the star formation age and timescale when considering the
stacked spectrum across the entire arc.

This application demonstrates the usefulness of collaborative nested
sampling in a realistic dataset. Physical parameters were extracted
while exploiting that spatial neighbours have similar physical properties.
However, no assumption about smoothness or neighbourhood was made,
i.e., the constraints at each pixel are independent. Also note that
in every pixel, a posterior distribution over the parameters is derived.
Here, the collaborative nested sampling analysis of 4223 fibers required
14.4 million likelihood evaluations (140h). This corresponds to a
quadrupling of the efficiency compared to analysing only 100 fibers,
which required 2.8 million likelihood evaluations (14.9h).

\section{Discussion\label{sec:Discussion}}

Collaborative nested sampling is a scalable algorithm suitable for
analysing massive data sets with arbitrarily complex physical models
and complex, inhomogeneous noise properties. The algorithm brings
to the Big Data regime parameter estimation with uncertainties, classification
of objects and distinction between physical processes.

The key insight in this work is to take advantage of a property specific
to nested sampling: The sampling regions can look similar across similar
data sets, and rejection sampling from the union of contours is permitted\footnote{As in classic nested sampling, the volume shrinkage estimates are
valid on average. Multiple runs can test whether this leads to additional
scatter in the integral estimate. In practice, single runs already
give correct uncertainties for many problems.}. \emph{Collaborative nested sampling} reduces the number of unique
model evaluations, in particular at the beginning of the nested sampling
run. The same approach cannot be followed with Markov Chain proposals:
There, the proposal depends on the current point, and deviating acceptances
prohibit a later joint proposal. 

Compared to embarrassingly parallel analyses, collaborative nested
sampling excels in specific types of problems, which have many uncertain
data sets of similar structure into which a slow physical model is
predicting. The algorithm has some overhead related to the management
of live points, in particularly to determine the unique set of live
points across a dynamically selected subgroup of data sets. The memory
usage also grows if big data sets have to be held in the same machine.
If only chunks of $N$ are managable, the analyses can be split into
such sizes and analysed in parallel across multiple machines. In that
case, one can take advantage of the scaling of the algorithm until
$N$.

The algorithm can be applied immediately to any existing large data
sets. Compared to other Big Data analysis approaches, nested sampling
supports model comparison and yields full probability distributions
on arbitrary models, allowing the exploration of degenerate fit solutions.
Furthermore, the instrument response can be modelled out and separated
from the process of interest. To give one application example, \emph{eROSITA}
\citep{Predehl2014eRosita} requires the source classification and
characterisation of 3 million point sources in its all-sky X-ray survey
\citep{Kolodzig2012}. The position-dependent detector response and
non-Gaussianity of count data make standard machine learning approaches
difficult to apply.

Even in the analysis of single objects the presented algorithm can
help. One might test the correctness of selecting a more complex model,
e.g., based on Bayes factors, as in the toy example presented. Large
Monte Carlo simulations of a null hypothesis model can be quickly
analysed with the presented method, with a model evaluation cost that
is essentially independent of the number of generated data sets. \r{Going
further, approaches to validate models and Bayesian inference \citep[e.g.,][]{Talts2018}
over the entire parameter space can be sped up.}

\section{Acknowledgements}

I thank Surangkhana Rukdee and Frederik Beaujean for reading the manuscript,
and Franz E. Bauer for help with MUSE data. This work made use of
the \textsc{nestle} free software implementation\footnote{\href{https://github.com/kbarbary/nestle/}{https://github.com/kbarbary/nestle/}}
of the \textsc{MultiNest} algorithm and the matplotlib plotting library
\citep{matplotlib}. I thank the two anonymous referees for helpful
comments and suggestions.

I acknowledge support from the CONICYT-Chile grants Basal-CATA PFB-06/2007,
FONDECYT Postdoctorados 3160439 and the Ministry of Economy, Development,
and Tourism's Millennium Science Initiative through grant IC120009,
awarded to The Millennium Institute of Astrophysics, MAS. This research
was supported by the DFG cluster of excellence ``Origin and Structure
of the Universe''.

\bibliographystyle{spmpsci}
\bibliography{agn}

\end{document}